\documentclass{PoS}

\title{
{\vspace{-25mm}\normalsize\hfill{\small BI-TP 2010/41}}\\[20mm]
Effective Polyakov-loop theory for pure Yang-Mills from strong coupling
expansion}

\ShortTitle{Effective Polyakov-loop theory}

\author{\speaker{Jens Langelage}\\
		Fakult\"at f\"ur Physik, Universit\"at Bielefeld, \\
		33501 Bielefeld, Germany \\
        E-mail: \email{jlang@physik.uni-bielefeld.de}}

\author{{Stefano Lottini$^*$}, Owe Philipsen\\
        Institut f\"ur Theoretische Physik - Johann Wolfgang Goethe-Universit\"at\\
        Max-von-Laue-Str. 1, 60438 Frankfurt am Main, Germany \\
        E-mail: \email{lottini, philipsen @th.physik.uni-frankfurt.de}}

\abstract{
Lattice Yang-Mills theories at finite temperature can be mapped onto effective 3d spin systems,
thus facilitating their numerical investigation. Using strong-coupling expansions
we derive effective actions for Polyakov loops in the $SU(2)$ and $SU(3)$ cases
and investigate the effect of higher order corrections.
Once a formulation is obtained which allows for Monte Carlo analysis,
the nature of the phase transition in both classes of models is investigated numerically,
and the results are then used to predict -- with an accuracy within a few percent --
the deconfinement point in the original 4d Yang-Mills pure gauge theories,
for a series of values of $N_\tau$ at once.
}

\FullConference{The XXVIII International Symposium on Lattice Field Theory\\
                 June 14-19,2010\\
                 Villasimius, Sardinia Italy}

\usepackage{graphicx}
\usepackage{amssymb,amsmath}
\usepackage[utf8]{inputenc}

\newcommand{\eqa}{\begin{eqnarray}}
\newcommand{\qea}{\end{eqnarray}}
\newcommand{\eq}{\begin{equation}}
\newcommand{\qe}{\end{equation}}

\newcommand{\de}{\mathrm{d}\hspace{0.08em}}

\newcommand{\tr}{\mathrm{Tr}}

\newcommand{\Real}{{\rm Re}}
\newcommand{\avg}[1]{\langle #1 \rangle}
\newcommand{\bigavg}[1]{\Big\langle #1 \Big\rangle}

\begin{document}

\makeatletter
\setbox\@firstaubox\hbox{\small J.~Langelage, S.~Lottini}
\makeatother

\section{Introduction}

In the framework of non-abelian gauge theories at finite temperature, several effective descriptions
have been pursued in order to overcome the infrared problems 
\cite{linde} connected with perturbative approaches
to the fundamental theory.
A rather successful technique is dimensional reduction \cite{dr1,dr2}. Thanks to the presence
of different energy scales, 
induced by the finite temperature dynamics 
of the original $(3+1)$-dimensional
theory, an integration over the hard modes leads to a 3d effective description which can then be 
solved in a non-perturbative way (e.~g.~by Monte Carlo integration).

In the case of QCD, this technique loses its validity in the confined phase; however, one would want to
devise effective methods to study the vicinity of the deconfinement transition: this is not a completely 
trivial task since the standard perturbative dimensional reduction does not retain the $Z(N)$ symmetry of the original Yang-Mills 
theory \cite{drqcd}.
One can then follow a different strategy, namely writing down a general theory respecting the desired 
symmetry and then fixing the (many) couplings by matching with particular observables \cite{vy,rob}. 
While for $SU(2)$ the phase transition is captured correctly by such approaches \cite{fk,dumitru_smith},
for the physically relevant $SU(3)$ gauge theory a satisfactory fixing of all couplings is still an open
issue.

A different way to pin down a 3d effective theory is to employ lattice strong coupling expansions.
This idea, first considered in \cite{Svetitsky:1982gs}, has been pursued by various authors
\cite{Polonyi:1982wz,Green:1983sd,Gocksch:1984yk,Gross:1984wb,Heinzl:2005xv,Wozar:2007tz}
and leads to theories with Polyakov loops as fundamental degrees of freedom.
The contribution from spatial plaquettes was often neglected, a simplification which preserves
the universal behaviour of the theory; in \cite{Billo:1996wv}, instead, they were explicitly
taken into account. Recent developments including staggered fermions can be found in \cite{Nakano:2010bg}.

The models proposed here systematically extend this approach 
by providing series for the effective couplings
up to a certain order and are
thus valid beyond the spatial strong coupling limit.
As is to be expected from strong-coupling expansions, our results
will have a finite radius of convergence, which is supposed to coincide with
the deconfinement transition: in this sense, our effective formulation
is complementary to weak coupling approaches.
The effective actions we propose are subsequently studied by means of Monte Carlo
integration, and the results are shown to lead to the correct order of the 
transition as well as to good estimates of the deconfinement point.

\section{Derivation of the effective theory}
\label{sec:deriv}

\subsection{General strategy and $SU(2)$}

Consider the partition function of a $(3+1)$-dimensional lattice gauge
field theory at finite temperature $\left(T=\frac{1}{aN_\tau}\right)$ with 
gauge group $SU(N)$ and Wilson's gauge action
\begin{eqnarray}
Z=\int\left[dU_0\right]\left[dU_i\right]\exp\left[\frac{\beta}{2N}\sum_p
\left(\mathrm{tr}\;U_p+\mathrm{tr}\;U_p^{\dagger}\right)\right], \quad \beta=\frac{2N}{g^2}\;.
\label{eq:original_gaugetheory}
\end{eqnarray}
Finite temperature and the bosonic nature of the degrees of freedom imply the use of 
periodic boundary conditions in the time direction.

In order to arrive at an effective three-dimensional theory, we integrate
out the spatial degrees of freedom and get schematically \cite{Gross:1984wb}
\begin{eqnarray}
Z&=&\int\left[dU_0\right]\exp\left[-S_\mathrm{eff}\right]\;;\nonumber\\
-S_\mathrm{eff}&=&\ln\int\left[dU_i\right]\exp\left[\frac{\beta}{2N}\sum_p
\left(\mathrm{tr}\;U_p+\mathrm{tr}\;U_p^{\dagger}\right)\right]
\equiv\lambda_1S_1+\lambda_2S_2+\ldots\;.
\label{eq_seff}
\end{eqnarray}
We expand around $\beta=0$ and arrange the
effective couplings $\lambda_n=\lambda_n(\beta,N_\tau)$ 
in increasing order in $\beta$ of their leading terms.
Thus, the $\lambda_n$ become less 
important the higher $n$. 
As we shall see, the interaction terms
$S_n$ depend only on Polyakov loops
\begin{eqnarray}
 L_j\equiv\mathrm{tr}\;W_j\equiv
\mathrm{tr}\;\prod_{\tau=1}^{N_\tau}U_0(\vec{x}_j,\tau)\;.
\end{eqnarray}
With sufficiently accurate knowledge of the relations 
$\lambda_n(\beta,N_\tau)$, we are able to convert the 
couplings of the three-dimensional theory to those of the full theory.
Determining the 
critical parameters $\lambda_{n,c}$ 
of the effective theory then gives a whole
array of critical $\beta_c(N_\tau)$ for - in principle - all $N_\tau$. 
In the following we calculate strong coupling, i.e.~small $\beta$, expansions  
of the leading $\lambda_{n}$.

Since the calculations are quite similar for different numbers of colours, we 
now specialise
our derivation to the simpler case of $SU(2)$ and later provide the necessary 
changes for $SU(3)$. For more details see \cite{Langelage:2010yr}.
Using the character expansion as described e.g.~in \cite{Montvay:1994cy,
Drouffe:1983fv}, the 
effective action according to Eq.~(\ref{eq_seff}) can be written as
\begin{eqnarray}
-S_\mathrm{eff}=\ln\int\left[dU_i\right]\prod_p\left[1+\sum_{r\neq0}d_ra_r(\beta)
\chi_r(U_p)\right]\;,
\label{eq_char}
\end{eqnarray}
where the sum extends over all irreducible representations $r$ with dimension 
$d_r$ and character $\chi_r$. The expansion coefficients $a_r(\beta)$ are 
accurately known \cite{Montvay:1994cy} and in the following we use $u\equiv a_f$  
as expansion parameter instead of $\beta$ for its better apparent convergence.
The logarithm in this definition allows us to use the method of moments
and cumulants \cite{Munster:1980wc}, and we get the following cluster expansion 
\begin{eqnarray}
-S_\mathrm{eff}&=&\sum_{C=(X_l^{n_l})}a(C)\prod_l\Phi\Big(X_l;\left\lbrace 
W_j\right\rbrace\Big)^{n_l}\;;\\
\Phi\Big(X_l;\left\lbrace 
W_j\right\rbrace\Big)&=&\int\left[dU_i\right]\prod_{p\in X_l}d_{r_p}a_{r_p}
\chi_{r_p}(U_p)\;,\nonumber
\label{eq_charexp}
\end{eqnarray}
where the combinatorial factor $a(C)$ equals 1 for a single polymer $X_i$ 
and $-1$ for two non-identical connected polymers. For clusters consisting of more than two 
polymers, $a(C)$ depends on how these polymers are connected. 
Our task is then to
group together all graphs yielding the same interaction terms up to some order 
in $\beta$, and this finally gives the strong coupling expansion 
of the corresponding effective coupling $\lambda_n$.

\subsection{Leading order effective action}

\begin{figure}
\hspace{-2cm}
\includegraphics[width=0.2\textwidth]{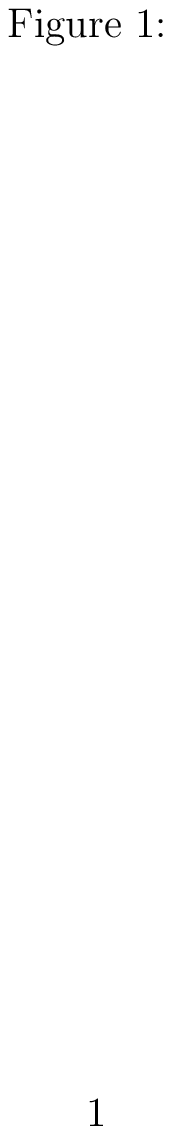}
\vspace{-2cm}
\caption{First graph with a nontrivial contribution 
after spatial integration for a lattice with temporal extent 
$N_\tau=4$. Four plaquettes in the fundamental representation lead to an 
interaction term involving two adjacent fundamental Polyakov loops 
$L_i$ and $L_j$.}
\label{fig_lo}
\end{figure}
The leading order result of the effective action has first been 
calculated in \cite{Polonyi:1982wz} and corresponds to a sequence 
of $N_\tau$ plaquettes that wind around the lattice in temporal direction, 
cf.~Fig.~\ref{fig_lo}..
Its contribution is given by:
\begin{eqnarray}
\lambda_1S_1=u^{N_\tau}\sum_{<ij>}L_iL_j\;.
\end{eqnarray}
Hence, to leading order the first coupling of the effective theory is 
$\lambda_1(u,N_\tau)=u^{N_\tau}$.

For additional terms of the series for $\lambda_1$, we 
can use most of the graphs that also appear in the strong coupling expansion of the 
Polyakov loop susceptibility \cite{Langelage:2009jb}. These corrections 
involve additional plaquettes, are hence of higher order in $u$ and we 
call these attached plaquettes decorations.
Carrying out the calculations, we get the following results through
order $u^{10}$ 
in the corrections relative to the leading order graph:
\begin{eqnarray}
\lambda_1(u,2)&=&u^2\exp\left[2\left(4u^4-8u^6+\frac{134}{3}u^8-\frac{49044}{405}u^{10}\right)\right]\;,\nonumber\\
\lambda_1(u,3)&=&u^3\exp\left[3\left(4u^4-4u^6+\frac{128}{3}u^8-\frac{36044}{405}u^{10}\right)\right]\;,\nonumber\\
\lambda_1(u,4)&=&u^4\exp\left[4\left(4u^4-4u^6+\frac{140}{3}u^8-\frac{37664}{405}u^{10}\right)\right]\;,\nonumber\\
\lambda_1(u,N_{\tau}\geq5)&=&u^{N_\tau}\exp\left[N_{\tau}\left(4u^4-4u^6+\frac{140}{3}u^8-\frac{36044}{405}u^{10}\right)\right]\;.
\label{eq_lambda}
\end{eqnarray}
For smaller $N_\tau$ some graphs do not contribute since the temporal 
extent of their 
decoration is $\geq N_\tau$ so that they do not fit into the lattice.

\subsection{Higher order terms}

There occur several types of higher order graphs: larger numbers of loops involved, Polyakov 
loops at distances larger than one and Polyakov loops in 
higher dimensional representations.
We begin by considering powers of the leading order term.
Inspection of higher order terms shows that one can arrange a subclass of 
these terms in the following manner
\begin{eqnarray}
\sum_{<ij>}\left(\lambda_1L_iL_j-\frac{\lambda_1^2}{2}L_i^2L_j^2+
\frac{\lambda_1^3}{3}L_i^3L_j^3-\ldots\right)=\sum_{<ij>}\ln\left(1+
\lambda_1L_iL_j\right)\;.
\label{eq_powers}
\end{eqnarray}
To see this, one 
calculates the corresponding graphs with $L_i^2L_j^2$ or $L_i^3L_j^3$, and the 
combinatorial factor $a(C)$ of Eq.~(\ref{eq_charexp}) gives the correct prefactors 
for the series to represent a logarithm. 

Next, let us consider couplings pertaining
to next-to-nearest neighbour interactions. These appear once additional
plaquettes are taken into account. Naively, the leading contribution should 
correspond to a planar graph with Polyakov loops at distance two. However, this
graph is precisely cancelled by the contribution of the nearest-neighbour graph squared
and its associated combinatorial factor $-1$.
The leading non-zero contribution therefore comes from L-shaped graphs and is given by
\begin{eqnarray}
\lambda_2(u,N_\tau)S_2=N_\tau(N_\tau-1)u^{2N_\tau+2}\sum_{\left[kl\right]}L_kL_l\;,
\end{eqnarray}
where we have two additional spatial plaquettes
and we sum over all pairs 
of loops with a diagonal distance of $\sqrt{2}a$, abbreviated by 
$\left[kl\right]$. With the same steps leading 
to Eq.~(\ref{eq_powers}), we finally arrive at the $SU(2)$ partition 
function
\begin{eqnarray}
Z=\int\left[dW\right]\prod_{<ij>}\left[1+\lambda_1L_iL_j\right]\prod_{\
\left[kl\right]}\left[1+\lambda_2L_kL_l\right]\;.
\label{eq:2coupling_su2}
\end{eqnarray}
Finally, we include some remarks about the Polyakov loops in higher 
dimensional representations. Consider, e.g., the adjoint Polyakov loop: the 
leading order term 
emerging from a strong coupling expansion is
\begin{eqnarray*}
\lambda_aS_a=v^{N_\tau}\sum_{<ij>}\chi_a(W_i)\chi_a(W_j)\;,\quad v=\frac{2}{3}u^2
+\frac{2}{9}u^4+\frac{16}{135} u^6+\ldots
\end{eqnarray*}
and hence $\lambda_a\sim u^{2N_\tau}$, which is formally of lower order than the coupling
$\lambda_2$. To next-to-leading order (valid for all $N_\tau\geq 2$) we have
\begin{equation}
\lambda_a=v^{N_\tau}\left( 1+N_\tau \frac{8}{3}\frac{u^6}{v}+\ldots \right).
\end{equation} 
Effects of higher representations have also been investigated in the literature 
\cite{Heinzl:2005xv,Wozar:2007tz,Dumitru2003:hp}.

\subsection{The effective action for $SU(3)$}

In the case of $SU(3)$ the same steps as for $SU(2)$ apply.
The only difference we have to keep in mind is that $SU(3)$ also has 
an anti-fundamental representation and consequently there is also
a complex conjugate Polyakov loop variable $L_i^{\ast}$.
Thus we get the one-coupling and two-coupling partition functions
\begin{eqnarray}
Z_1&=&\int\left[dW\right]\prod_{<ij>}\left[1+\lambda_1\left(L_iL_j^{\ast}+
L_i^{\ast}L_j\right)\right]\;,\label{eq:su3-onecoupling}\\
Z_2&=&\int\left[dW\right]\prod_{<ij>}\left[1+\lambda_1\left(L_iL_j^{\ast}+
L_i^{\ast}L_j\right)\right]\prod_{\left[kl\right]}\left[1+\lambda_2\left(L_kL_l^{\ast}+
L_k^{\ast}L_l\right)\right]\;.\label{eq:su3-twocoupling}
\end{eqnarray}
The effective coupling $\lambda_1(u,N_\tau)$ is obtained as 
(for this gauge group we consider only even values of $N_\tau$):
\begin{eqnarray}
\hspace{-0.4cm}\lambda_1(2,u)&=&u^2\exp\left[2\left(4u^4+12u^5-18u^6-36u^7\right.\right.\nonumber\\
&&\hspace*{2cm} \left.\left.
+\frac{219}{2}u^8+\frac{1791}{10}u^9+\frac{830517}{5120}u^{10}\right)\right]\;,\nonumber\\
\hspace{-0.4cm}\lambda_1(4,u)&=&u^4\exp\left[4\left(4u^4+12u^5-14u^6-36u^7\right. \right. \nonumber\\
&&\hspace*{2cm}\left.\left.
+\frac{295}{2}u^8+\frac{1851}{10}u^9+\frac{1035317}{5120}u^{10}\right)\right]\nonumber\;,\\
\hspace{-0.4cm}\lambda_1(N_{\tau}\geq6,u)&=&u^{N_\tau}\exp\left[N_{\tau}\left(4u^4+12u^5-14u^6-36u^7\right.\right.\nonumber\\
&&\hspace*{2.5cm}\left.\left.
+\frac{295}{2}u^8+\frac{1851}{10}u^9+\frac{1055797}{5120}u^{10}\right)\right]\;.
	\label{eq_lambda2}
\end{eqnarray}
For the first terms of the next-to-nearest neighbour coupling 
$\lambda_2(N_\tau,u)$ we find
\begin{eqnarray}
\lambda_2(2,u)&=&u^4\Big[2u^2+6u^4+31u^6\Big]\;,\nonumber\\
\lambda_2(4,u)&=&u^8\Big[12u^2+26u^4+364u^6\Big]\;,\nonumber\\
\lambda_2(6,u)&=&u^{12}\Big[30u^2+66u^4\Big]\;,\nonumber\\
\lambda_2(N_\tau\geq8,u)&=&u^{2N_\tau}\left[N_\tau(N_\tau-1)u^2\right]\;,
\end{eqnarray}
while the leading coupling of adjoint loops is (valid for $N_\tau\geq 2$)
\begin{equation}
\lambda_a=v^{N_\tau}\left( 1+N_\tau \frac{3}{2}\frac{u^6}{v}+\ldots \right),\quad
v=\frac{9}{8}u^2-\frac{9}{8}u^3+\frac{81}{32}u^4+\ldots
\label{adjlam}
\end{equation}

\section{Numerical simulation of the effective theories}
\label{sec:numerical}
\subsection{The one coupling model}

For the purpose of numerical simulations, a further simplification is achieved by using 
the trace of the Polyakov loops for the path integral measure as degrees of freedom 
(complex numbers, $|L_x|\leq 3$, instead of matrices), and rewrite the one-coupling partition function for 
$SU(3)$, Eq.~(\ref{eq:su3-onecoupling}),
\eq
	Z = \Big(\prod_x \int \de L_x\Big) e^{-S_\mathrm{eff}}
		\;;\;
	S_\mathrm{eff} = -\sum_{<ij>} \log(1+2\lambda_1 \Real L_i L^*_j)-
\sum_x V_x\;.
	\label{eq:su3_numerical_action}
\qe

The potential term $V_x$ is the Jacobian induced by the Haar measure of 
the original group integration; rotating the matrices to the diagonal form
$\mathrm{diag}(e^{i\theta},e^{i\phi},e^{-i(\theta+\phi)})$, with $|\theta|,|\phi| \leq \pi$,
we have \cite{gross_bartholomew_hochberg_1983}:
\eq
	V_x = \frac{1}{2}\log(27-18|L_x|^2+8\Real L_x^3-|L_x|^4)\;.
\qe
The integration measure actually used in our simulation then takes the form
\eq
	\int \de L_x e^{V_x} = \int_{-\pi}^{+\pi}\de\phi_x
		\int_{-\pi}^{+\pi}\de\theta_x e^{V_x}\;.
\qe
When working on the $SU(2)$ theory, $-2\leq L_x \leq +2$ is a real number 
and we simply have 
\eq
	\int_{-2}^{+2}\de L_x e^{V_x}\;,\;V_x = \frac{1}{2}\log(4-L_x^2)\;.
\qe

\subsection{A ``sign problem'' and its solution}
Our numerical approach will be a straightforward Metropolis local update algorithm;
however, the Boltzmann weights to consider are in the form
$\exp(\log(1+2\lambda_1\Real(L_iL^*_j)))$: for high enough couplings,
they can be also negative, thus spoiling the update technique
(the partition function being, overall, still positive).
In the $SU(2)$ case, the threshold coupling $\lambda^T=1/4$
is well beyond the phase transition, so that in practice
there is no problem around criticality, but the $SU(3)$ threshold
of $1/9$ is very close to the transition and a direct numerical investigation
of the model as in Eq.~(\ref{eq:su3_numerical_action}) is impossible.

Our approach to overcome this problem is the following: we Taylor-expand the 
logarithm in the effective action to some order $M$ in powers
of $q \equiv \lambda_1\Real L_i L^*_j$ (undoing 
the resummation as in Eq.~(\ref{eq_powers})), obtaining models
free of the problem:
	\eq
		S_\mathrm{eff}^{(M)} = -\sum_x V_x -\sum_{<ij>} \Big(
			2q -2q^2 +\frac{8}{3}q^3 - 4q^4+\frac{32}{5}q^5 - 
			\ldots - (-1)^M \frac{2^M}{M}q^M\Big)\;.
		\label{eq:taylor-m-expansion}
	\qe
In this way we can identify a critical point for each $M$ and
look for their convergence as $M\to\infty$. Also, we can compare
to the $SU(2)$ case where the $M=\infty$ value is directly calculable.

\subsection{Phase structure, critical coupling and finite size analysis}

\begin{figure}
\hspace*{-0.5cm}
\includegraphics[width=0.25\textwidth,angle=-90]{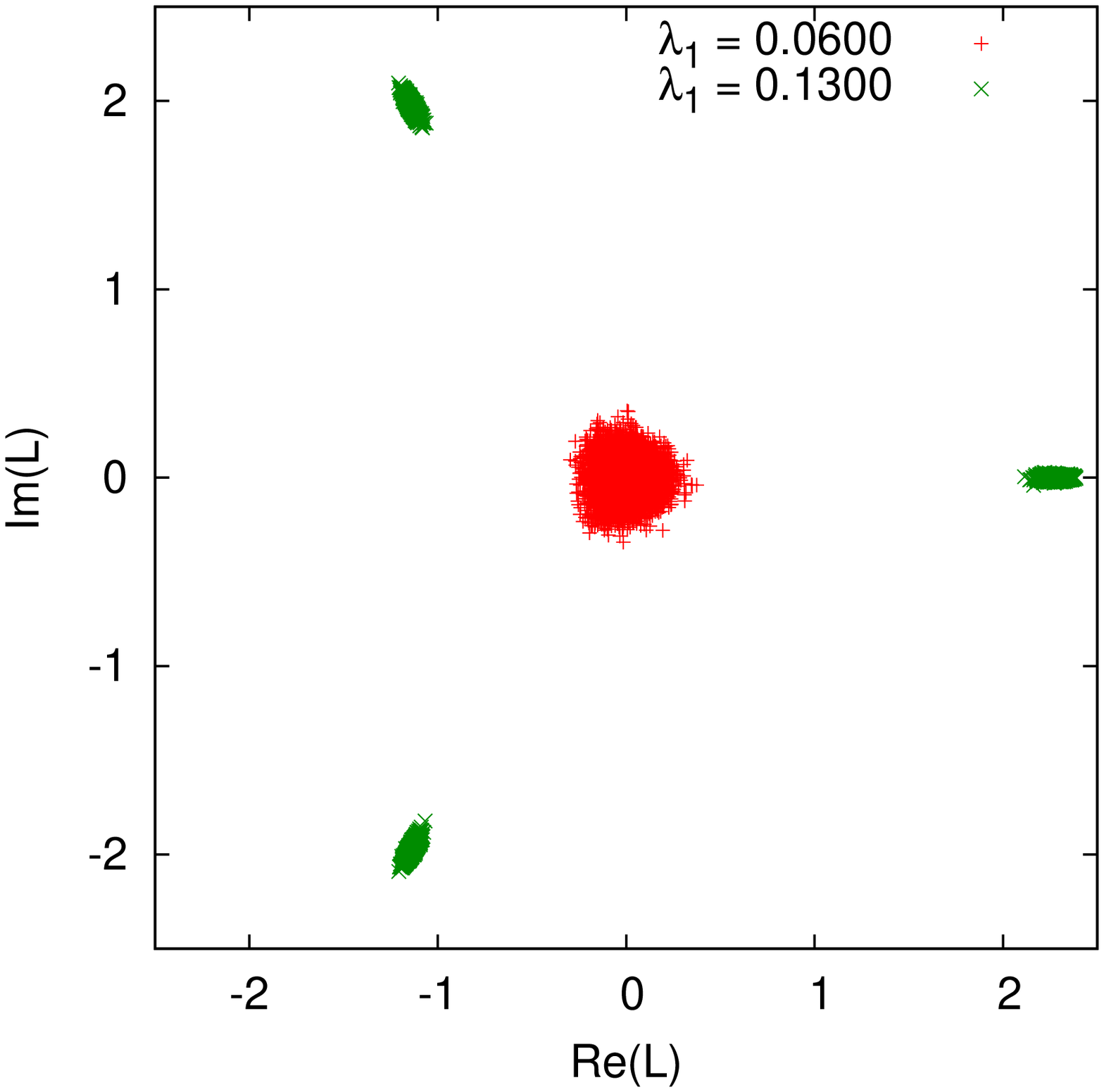}
\hspace*{-0.8cm}
\includegraphics[width=0.25\textwidth,angle=-90]{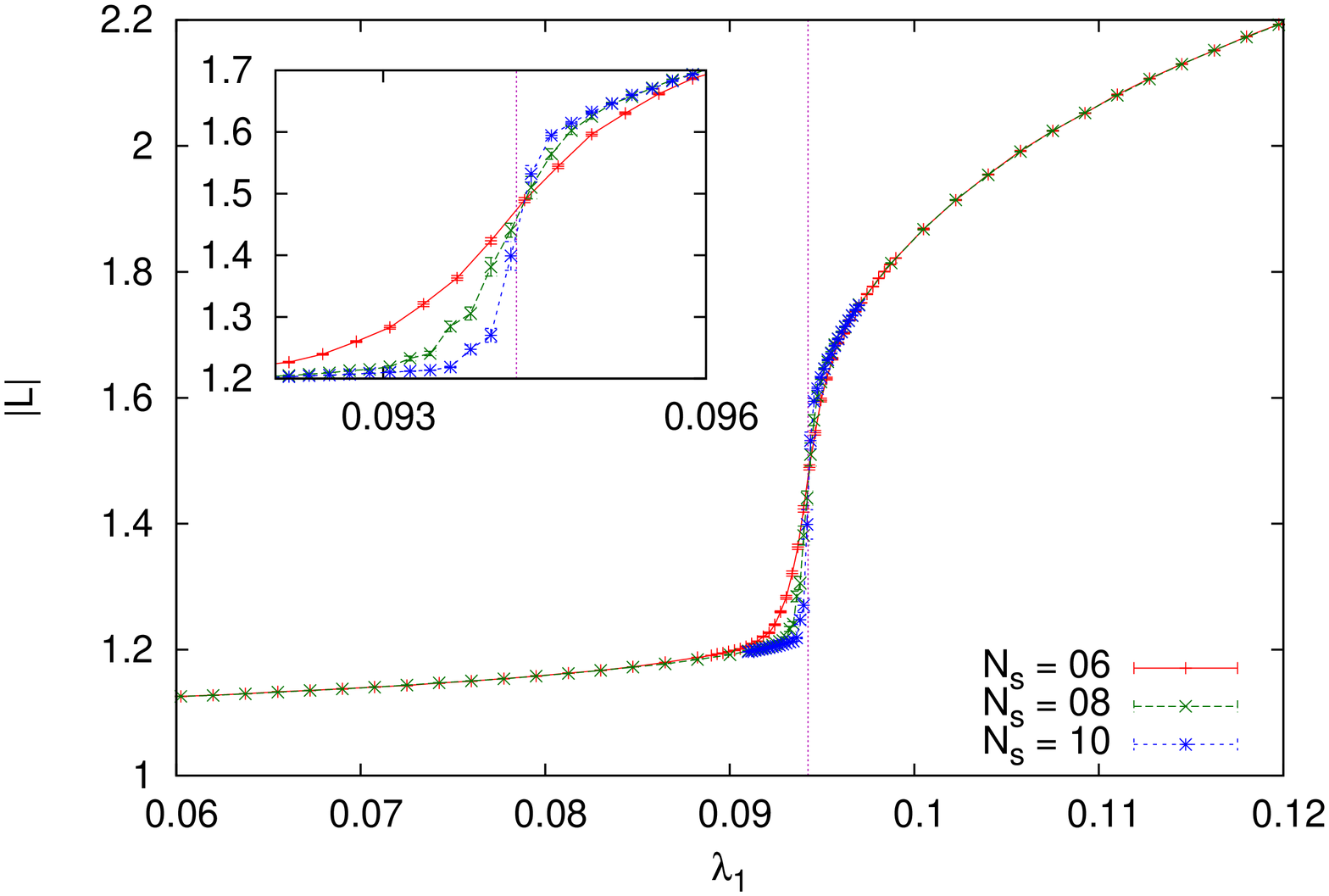}~
	\includegraphics[width=0.25\textwidth,angle=-90]{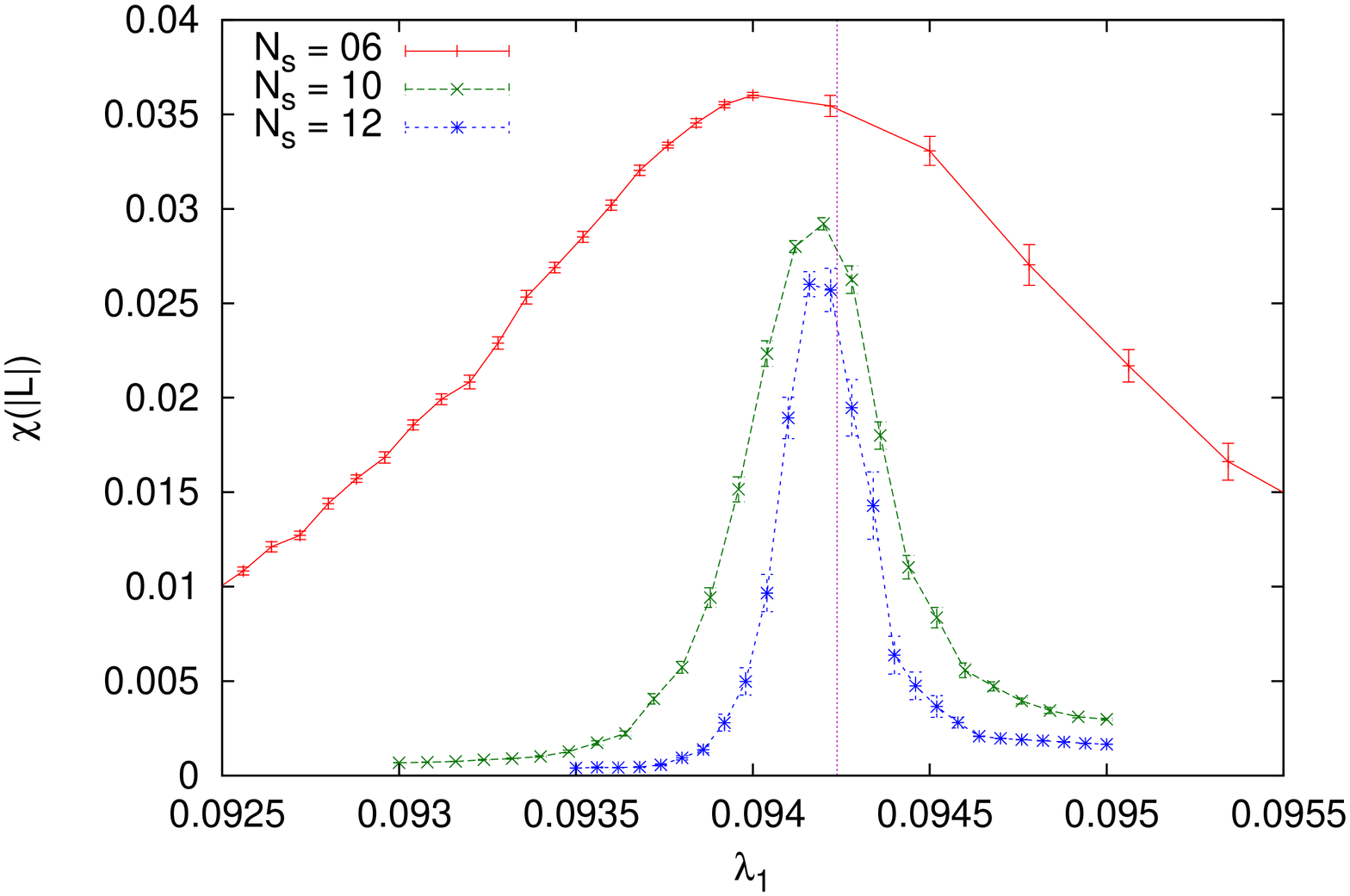}
\caption[]{Left: Distribution of $L$ for small and large $\lambda_1$ on a lattice with $N_s=6$ and $M=1$. Middle, Right: Expectation
value of $|L|$ and its susceptibility. The vertical line marks the infinite-volume transition.}
\label{fig:l}
\end{figure}

Our first task is to establish the phase structure of the effective theory, where we focus 
on the physically interesting case of $SU(3)$. Based on the global 
$Z(3)$ symmetry of the model, one expects spontaneous breaking of that symmetry 
for some critical value of the coupling $\lambda_{1,c}$. Fig.~\ref{fig:l} shows the behaviour
of the field variable $L$ as a function of $\lambda_1$. As expected from the 4d parent theory,
there is indeed a transition from a disordered or mixed phase, with values of $L$ 
scattering about zero, to an ordered phase at large coupling
where the three $Z(3)$-phases are populated separately.  In the thermodynamic limit, 
one of these vacua will be chosen and the symmetry is broken spontaneously,
$\langle L \rangle=0$ for $\lambda_1<\lambda_{1,c}$ and 
$\langle L \rangle \neq0$ for $\lambda_1>\lambda_{1,c}$. 
Correspondingly, the expectation value of $|L|$ rises abruptly at some critical coupling 
$\lambda_{1,c}$, as shown in Fig.~\ref{fig:l} (middle). On a finite size lattice, the phase
transition is smoothed out,  non-analyticities are approached gradually with growing volume, 
as the figure illustrates.

The critical coupling, $\lambda_{1,c}$, is located via finite-size scaling.
After identifying a pseudo-critical $\lambda_{1,c}(N_s)$ for a number of
finite systems, the relation
\eq
	\lambda_{1,c}(N_s) = \lambda_{1,c} + b N_s^{-1/\nu}\;
	\label{eq:fss-critical}
\qe
is used, with $\nu=1/3$ for the $SU(3)$ first-order transitions and, in the $SU(2)$ case,
the 3d Ising value $\nu = 0.63002$ \cite{ising_3d_index_nu}.
Numerically, we found satisfactory results with data produced in just a few 
days on a desktop PC.

For the definition of the pseudo-critical coupling, one can look at the energy
$E = -S_{\mathrm{eff}}/\lambda_1$ (neglecting the potential term)
or derived quantities, but in general, due to the
nonlinearity of $S_\mathrm{eff}$ in the coupling, we preferred to look at
the average modulus $|L|$; one can then define $\lambda_{1,c}$ as the
minimum/maximum of the associated Binder cumulant/susceptibility, which indeed 
featured a more robust scaling:
	\eq
		B(|L|) = 1 - \frac{\avg{|L|^4}}{3\avg{|L|^2}^2} \quad;\quad
		\chi(|L|) = \bigavg{\Big(|L|-\avg{|L|}\Big)^2}\;.
	\qe

\subsection{Critical coupling and order of the transition for $SU(3)$}
The truncated theories with $M=1,3,5$ were simulated on lattices with spatial sizes 
$N_s=6,8,10$ (plus $N_s=12$ for the $M=1$ theory). 
For each volume, $\sim 30$ values of the couplings are sampled 
by $\sim 10^6$ update sweeps each. Measurements 
were taken every $\sim 30$ updates.

\begin{figure}
\begin{center}
	\includegraphics[height=0.3\textwidth, angle=0]{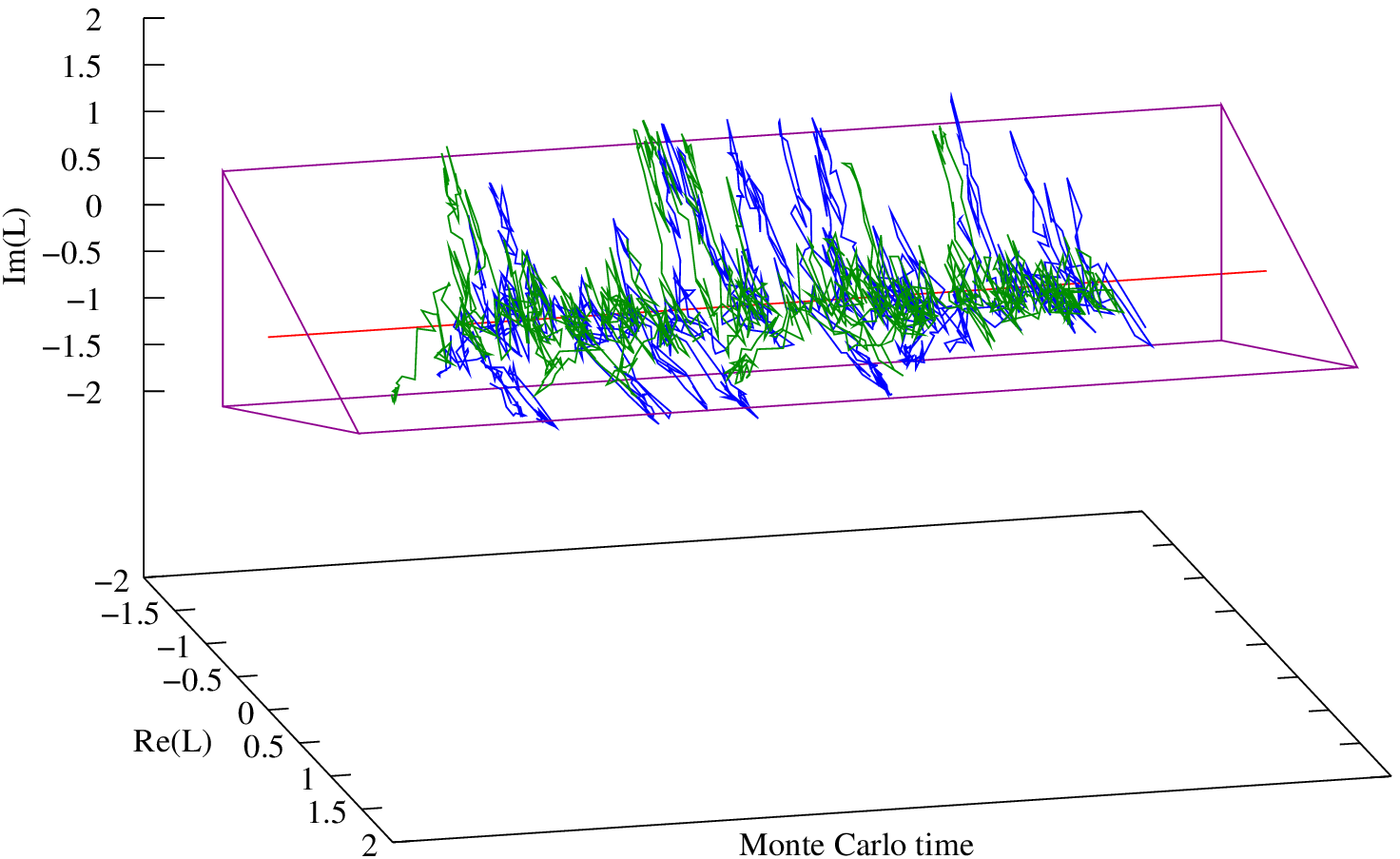}\hspace*{0.2cm}
	\includegraphics[height=0.3\textwidth, angle=0]{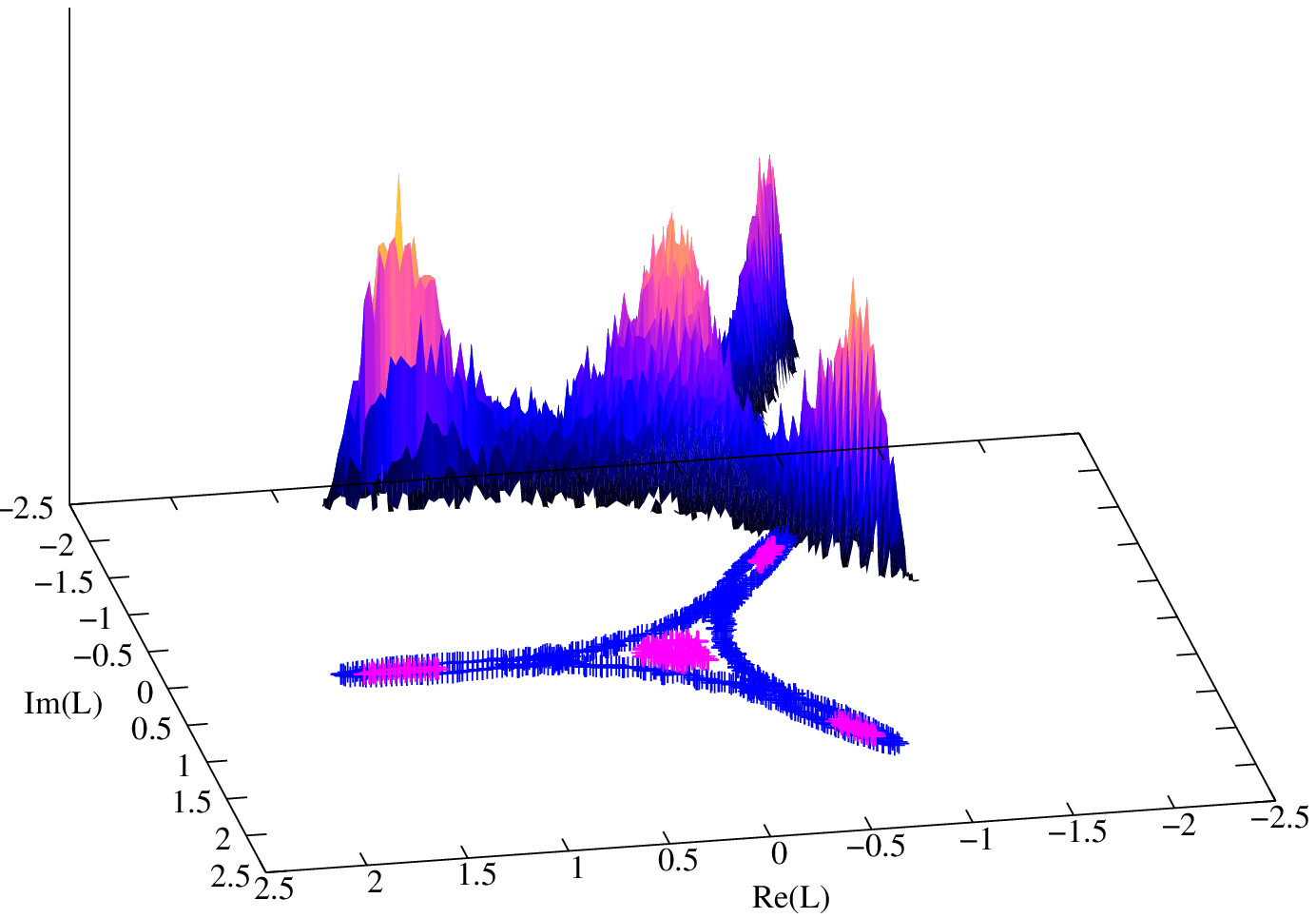}
 	\caption[]{Left: Behaviour of $L$ with Monte Carlo time for two $N_s=6$ trajectories in the $SU(3)$ $M=1$ theory
	with $\lambda_1=0.0935$.
		Right: Histogram for $L$, obtained from 60 such trajectories. The tunnelling between the central and the three
	broken-symmetry vacua is apparent.}
	\label{fig:tunnelling_etc}
\end{center}
\end{figure}

Regardless of the truncation order $M$, the $SU(3)$ theories display a first-order
transition; among the associated features, we found very long thermalisation times
$\propto \exp(c N_s^3)$ as is expected for tunnelling phenomena (Fig.~\ref{fig:tunnelling_etc}):
for instance, a system with size $N_s=16$ would require, around criticality,
$\sim 10^6$ update sweeps to thermalise.

\begin{figure}
\begin{center}
	\includegraphics[height=0.45\textwidth, angle=270]{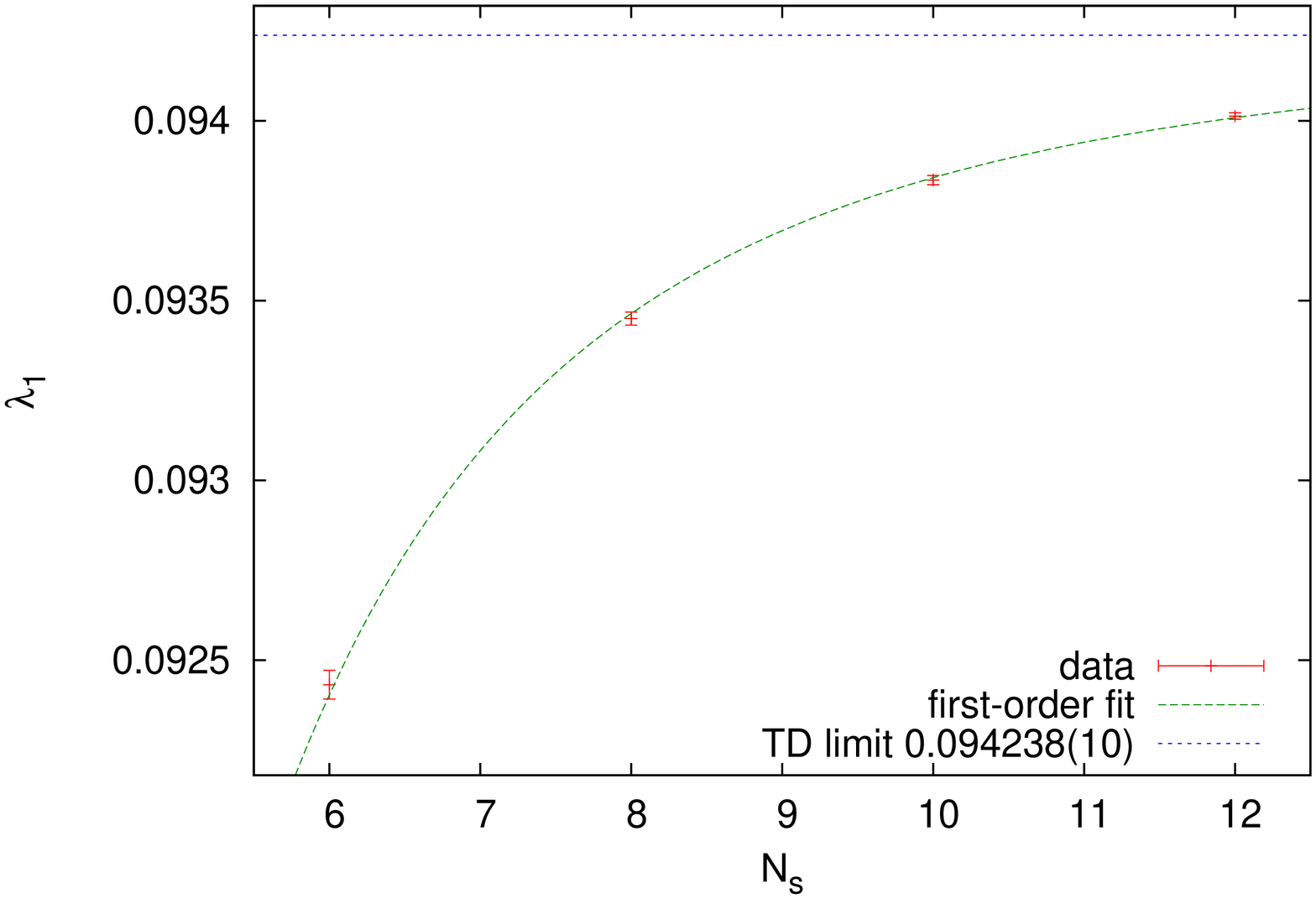}\hspace*{0.2cm}
	\includegraphics[height=0.45\textwidth, angle=270]{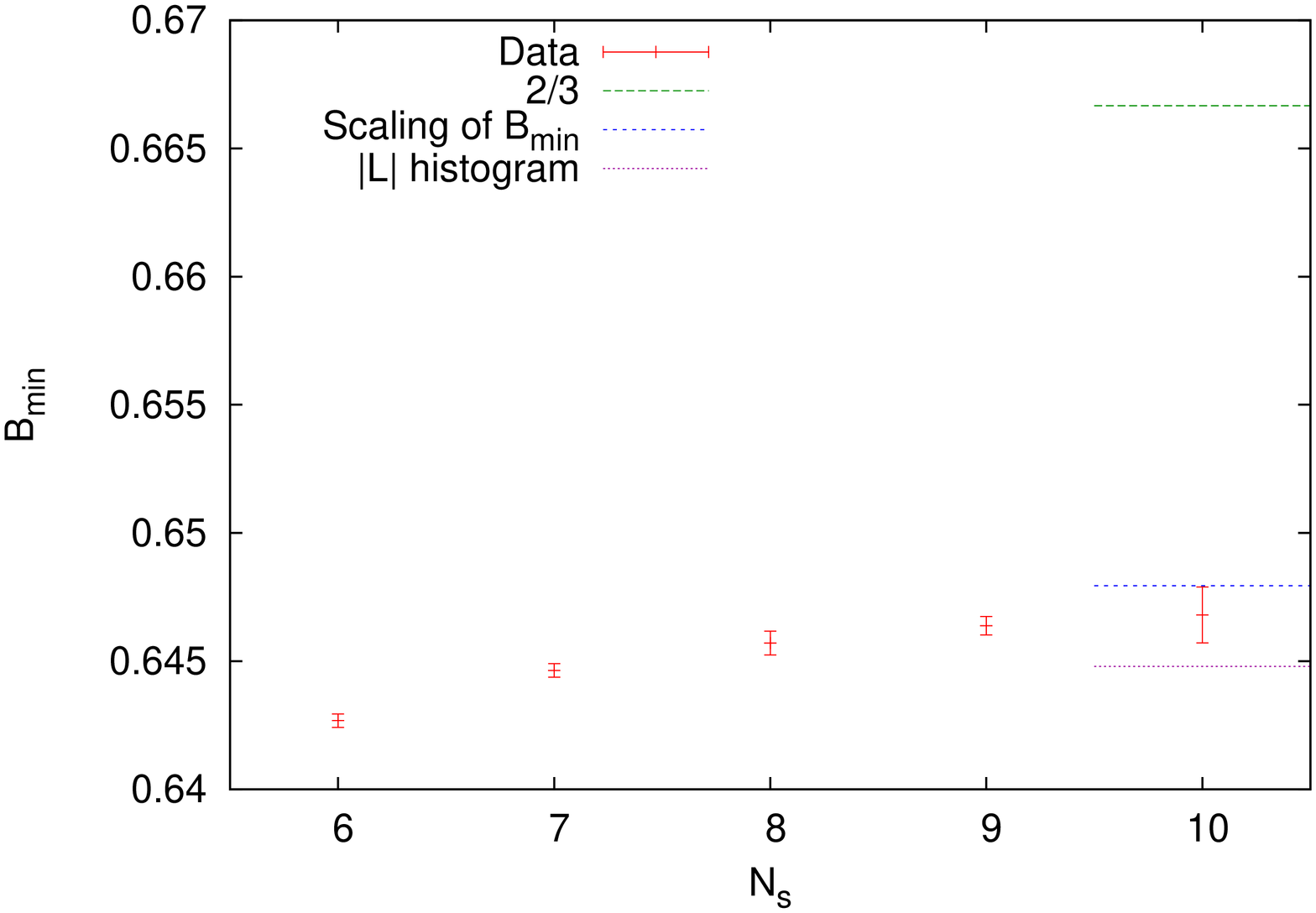}
 	\caption[]{Left: 
 		Position of the minimum of the Binder cumulant $B(E)$ for  $SU(3), M=1$, for different lattice sizes. 
		The horizontal 
		line is the thermodynamic limit resulting from the fit to Eq.~(\ref{eq:fss-critical}).
		Right: Behaviour of $B_\mathrm{min}(N_s)$, along with its thermodynamic limit obtained with the
		$\mathcal{O}(N_s^{-3})$ scaling law and the independent estimate $B_\infty$ from 
		the $|L|$ histogram. Also the second-order limit value $2/3$ is shown.}
	\label{fig:first-order-fss_m1_en}
\end{center}
\end{figure}

First we consider the model with $M=1$.
The first-order nature of the transition is established by fitting the pseudo-critical couplings to the
scaling law, Eq.~(\ref{eq:fss-critical}), with $\nu=1/3$, see Fig.~\ref{fig:first-order-fss_m1_en} (left).
The behaviour of the minimum $B_\mathrm{min}$ of $B(|L|)$ is a further confirmation; 
this quantity, as demonstrated in \cite{Binder1_lee_kosterlitz, Binder2_billoire_neuhaus_berg},
scales as $B_\mathrm{min}(N_s) = B_\infty + B^{(2)} N_s^{-3} + \mathcal{O}(N_s^{-6})$, with a thermodynamic
limit which is smaller than the second-order value $2/3$,
\eq
	B_\infty = \frac{2}{3} - \frac{1}{12}\Big( \frac{|L|_1}{|L|_2} - \frac{|L|_2}{|L|_1} \Big)^2 \;,
	\label{bestimate}
\qe
with $|L|_1$ and $|L|_2$ the two local maxima of the $|L|$ double-peaked histogram.
A direct comparison between the results for $B_\infty$ from
scaling analysis and from the location of $|L|_i$ shows an agreement within two standard deviations,
the residual discrepancy being probably due to neglecting higher-order $N_s^{-6}$ corrections.

\begin{figure}
\begin{center}
	\includegraphics[height=0.45\textwidth, angle=270]{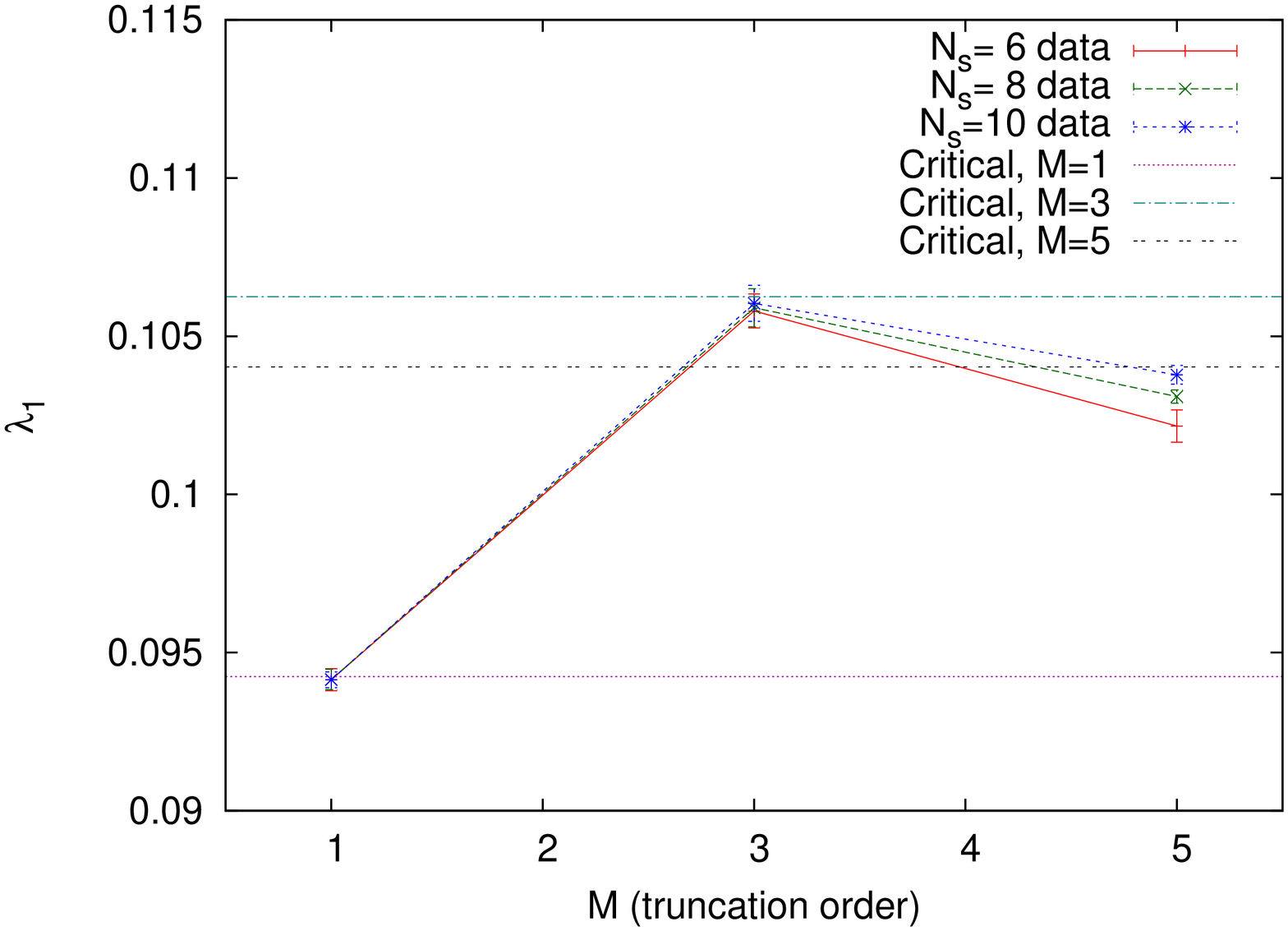}\hspace*{0.2cm}
	\includegraphics[height=0.45\textwidth, angle=270]{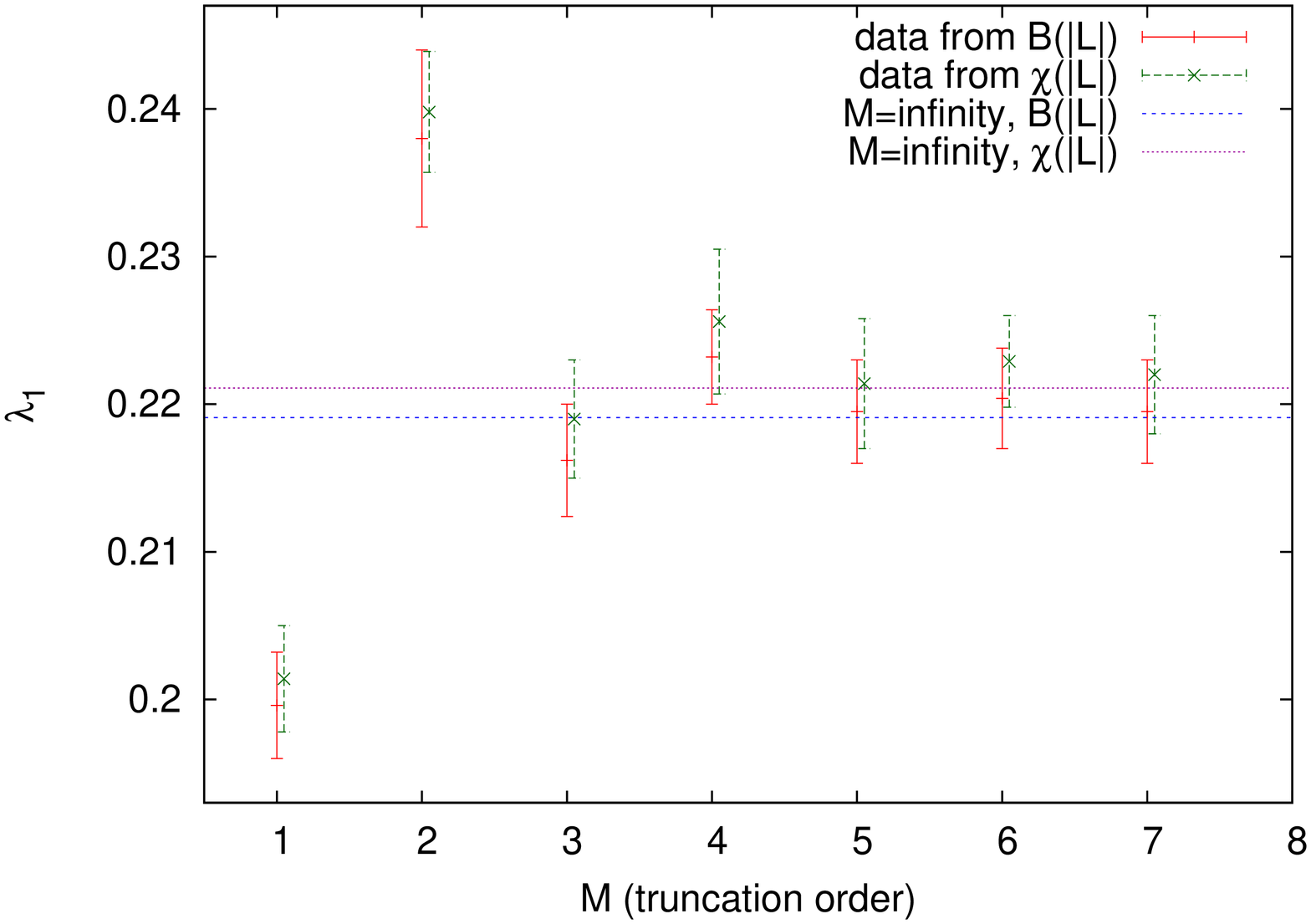}
	\caption{Truncation-dependence of the critical points. Left: in the $SU(3)$ case, the data points
	refer to three system sizes, while the lines mark the extrapolated critical points;
	the latter are found at $\lambda_{1,c} = 0.094238(10), 0.10635(11), 0.10403(28)$
	for $M=1,3,5$ respectively.
	Right: for $SU(2)$, pseudo-critical points for $N_s=8$ obtained from $B(|L|)$ and $\chi(|L|)$
	for a variety of truncations (data points) and compared with the untruncated $M=\infty$
	values (lines).}
	\label{fig:su_2-3_M}
\end{center}
\end{figure}

In the next step we need to investigate the behaviour of the models with higher $M$.
Again we observe first order transitions,
which become 
sharper with increasing $M$. Moreover, finite-size effects are 
stronger for higher $M$, Fig.~\ref{fig:su_2-3_M} (left). The critical 
couplings identified for the $M=1,3,5$ effective theories in the thermodynamic limit
are also quoted there.
Judging from these three values, the series seems to be rapidly converging, with only $\sim 3\%$ 
difference between $M=3,5$. The residual difference between this 
estimate and the $M=\infty$ critical coupling is completely subdominant 
compared to the other systematic errors contributing to 
the final results. Also, the direct comparison with the $SU(2)$ case below, 
where the $M=\infty$ data are directly available, supports 
a rapid convergence, Fig.~\ref{fig:su_2-3_M} (right).

\subsection{Critical coupling and order of the transition for $SU(2)$}

In this family of theories the transition is second-order; with much less
relaxation problems (e.g. 4000 steps for $N_s=16$),
larger lattices (up to $N_s=28$) were available. With the same approach as for $SU(3)$,
the nature of the transition was confirmed by: \textit{(a)} $\lambda_{1,c}(N_s)$ scaling with the
3d Ising critical index, \textit{(b)} Binder cumulant analysis approaching $2/3$ for large systems,
and \textit{(c)} $|L|$ histogram inspection, where a single peak continuously moves to the right as
the coupling is increased.
All inspected values of $M$ yielded the same features. Moreover,
here a direct comparison with the $M=\infty$ untruncated model is possible, and shows that
a rapid convergence is indeed realised (Fig.~\ref{fig:su_2-3_M}, right); in particular,
we found
\eq
	\lambda_{1,c}(M=1)=0.195374(42)\;\;;\;\;\lambda_{1,c}(M=\infty)=0.21423(70)\;,
\qe
which indicates quite small systematic deviations due to choosing
one particular truncation.
 
\subsection{Two-coupling models for $SU(3)$}

In this section we study the influence of including a second coupling. We consider two possibilities: the
first one is switching on the interaction between next-to-nearest neighbours. The $SU(3)$ version of
Eq.~(\ref{eq:2coupling_su2}) reads:
\eq
	Z = \Big(\prod_x \int \de L_x\Big) \prod_{<ij>}(1+2\lambda_1 \Real L_i L^*_j) \prod_{[kl]}(1+2\lambda_2 \Real L_k L^*_l)
	e^{\sum_x V_x}\;.
	\label{eq:2coupling_su3}
\qe
We remark that now there are two terms suffering from the above-mentioned sign problem: a truncated expansion is then 
needed in both, and the two truncation parameters $(M_1,M_2)$ should be chosen in a consistent way, for all $N_\tau$,
with respect to the power in $u$ we want to keep. We adopted the choice $(3,1)$ after checking numerically that higher
values of $M_2$ give negligible differences in the results.

In the other model, we allow the nearest neighbours to interact also in the adjoint
representation as described before. The partition function in this case (with the adjoint part
already truncated at $M_2=1$) is given by
\eq
        Z = \Big(\prod_x \int \de L_x\Big) \prod_{<ij>}(1+2\lambda_1 \Real L_i L^*_j)
  \prod_{<ij>}e^{\lambda_a (\tr^{(a)}W_i)(\tr^{(a)}W_j)}
        e^{\sum_x V_x}\;,
\qe
with the adjoint trace $\tr^{(a)}W = |\tr W|^2-1$. Also in this case, the truncation $(3,1)$ was employed.

In these two-dimensional parameter spaces, there is a critical line separating the symmetric
and the broken phases. However, for a given $N_\tau$, only a one-dimensional manifold in this
space represents the image of the original gauge theory, since both couplings are functions of
the sole $u$. The strategy was then to identify the shape of the critical
line and find, for each temporal lattice extent, the intersection with the curve enforcing that
particular value of $N_\tau$.

In both models, the critical lines were found by interpolation after locating 11 critical points at
as many fixed values of the second coupling; it turned out that a linear parametrisation was good enough
in describing them (within our precision, finite-size effects were practically invisible):
\eqa
	\lambda_{1,c} &=& a+ b \lambda_2 \quad \mbox{with}\quad a = \phantom{+}0.10628(\phantom{1}8),\; b = -1.891(\phantom{2}4)\;.\\
	\lambda_{1,c} &=& a+ b \lambda_a \quad \mbox{with}\quad a = \phantom{+}0.10637(15),\; b = -1.422(22)\;.
\qea
The value of $a$ was always, as expected, compatible with the estimate for the critical point
of the $M=3$ one-coupling theory.

\begin{figure}
\begin{center}
	\includegraphics[height=6.2cm,angle=270]{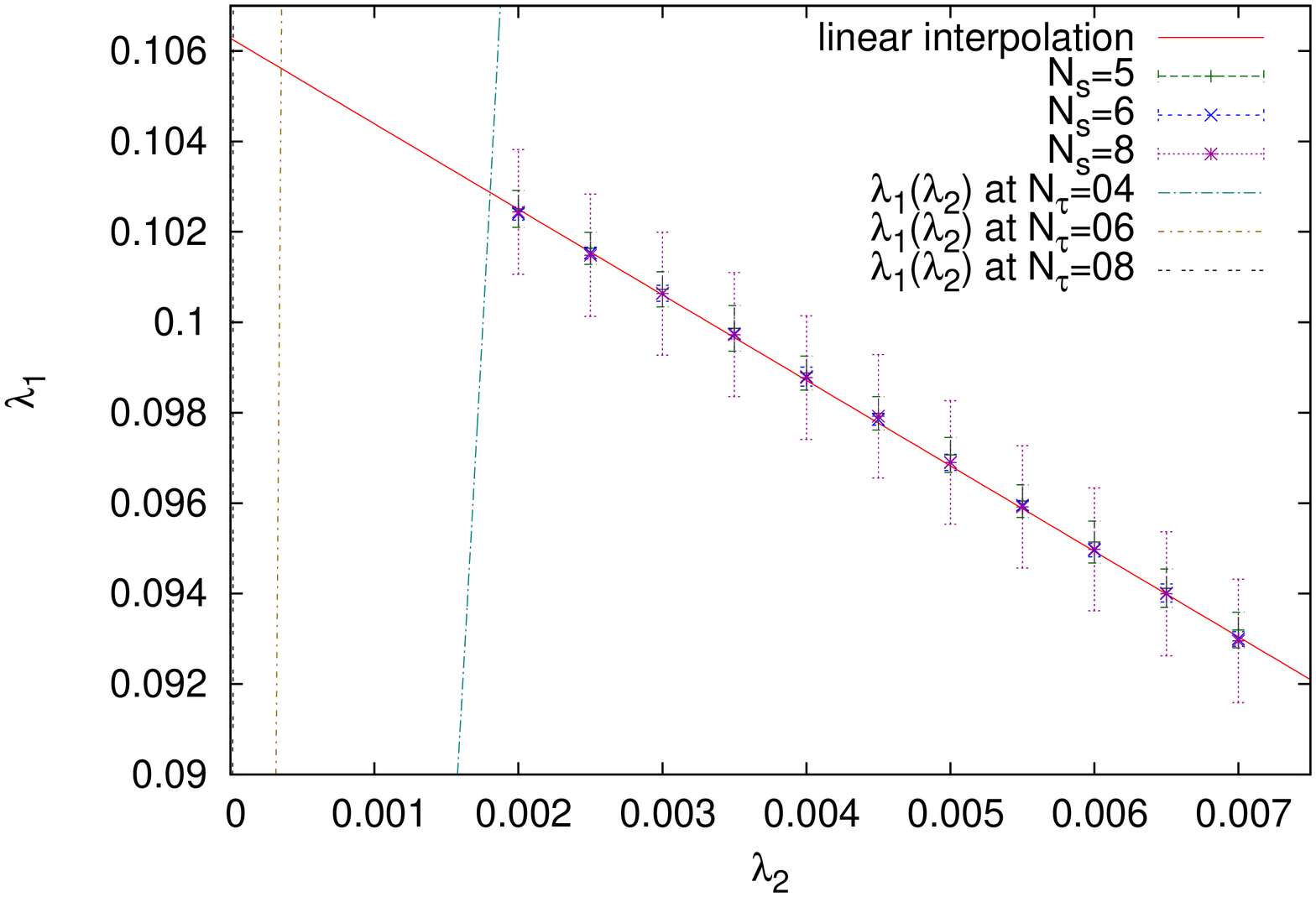}
	\includegraphics[height=6.2cm,angle=270]{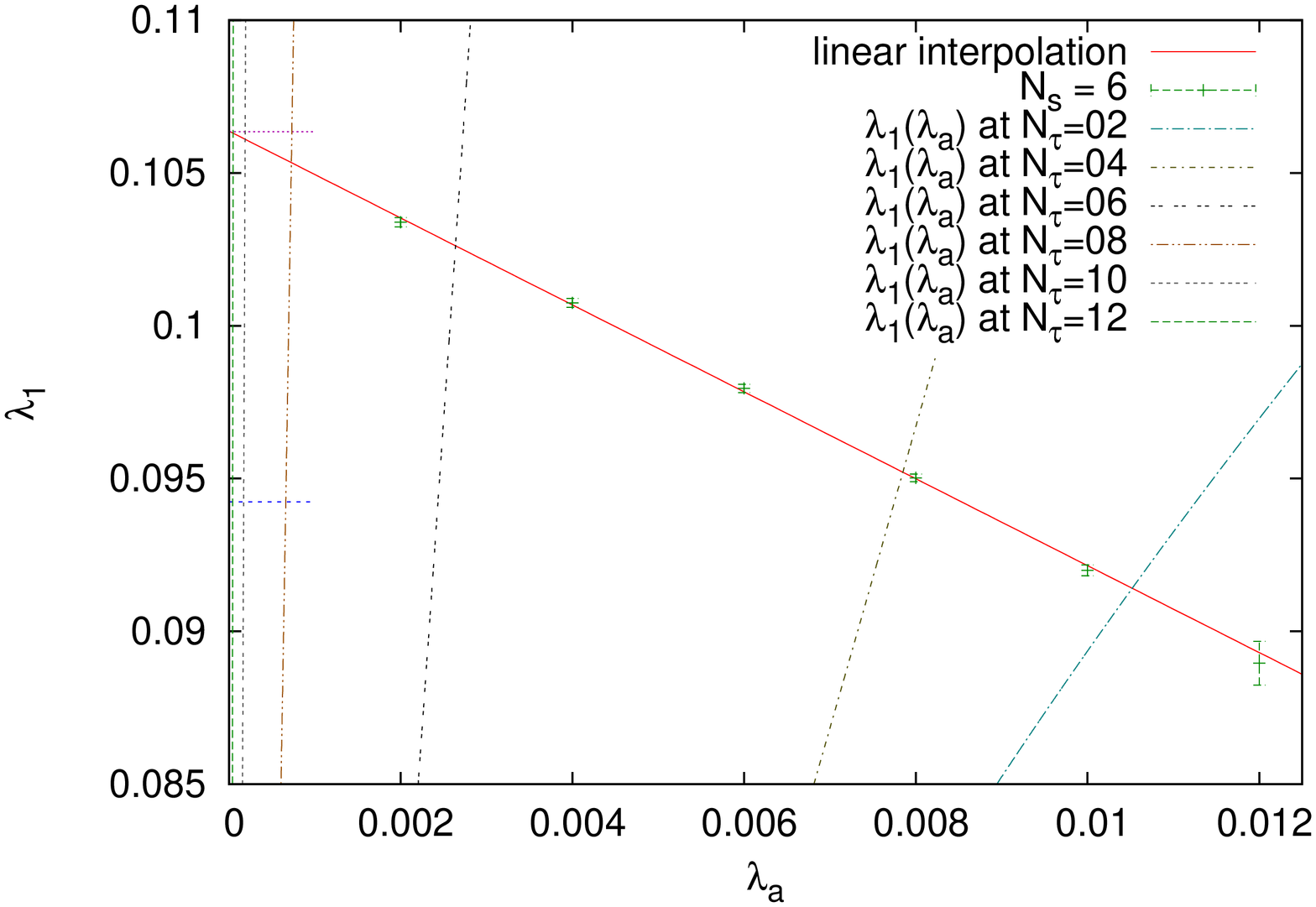}
 	\caption[]{Critical line in the two-coupling space, determined from $\chi(|L|)$.
	Dashed lines give the parameter space representing a 4d theory with fixed $N_\tau$.
	Left: $(\lambda_1,\lambda_2)$. Right: $(\lambda_1,\lambda_a)$.}
	\label{fig:critline-2coupling}
\end{center}
\end{figure}

By plotting these critical lines and the family of curves coming from requiring a given $N_\tau$, one sees
that the latter accumulate towards vanishing second-coupling as $N_\tau$ increases
(Fig.~\ref{fig:critline-2coupling}): this implies that the effect of including those interactions is less and
less important at finer lattice spacings: only at very low values of temporal extent does the inclusion
of a second coupling make any visible difference.

\section{Mapping back to 4d Yang-Mills}
\label{sec:results}

Having established the critical couplings for our effective theories and tested their 
reliability, we are now ready to map them back to the original thermal Yang-Mills theories
by using Eqs.~(\ref{eq_lambda}, \ref{eq_lambda2}).
In Tables \ref{tab:su2_betas}, \ref{tab:su3_betas} we collect the values for 
the critical gauge couplings, $\beta_c$, obtained in this way from the effective
theories and compare them to the values obtained from
simulations of the full 4d theories for $SU(2), SU(3)$, respectively. 

\begin{figure}
\begin{center}
	\includegraphics[height=6.2cm,angle=270]{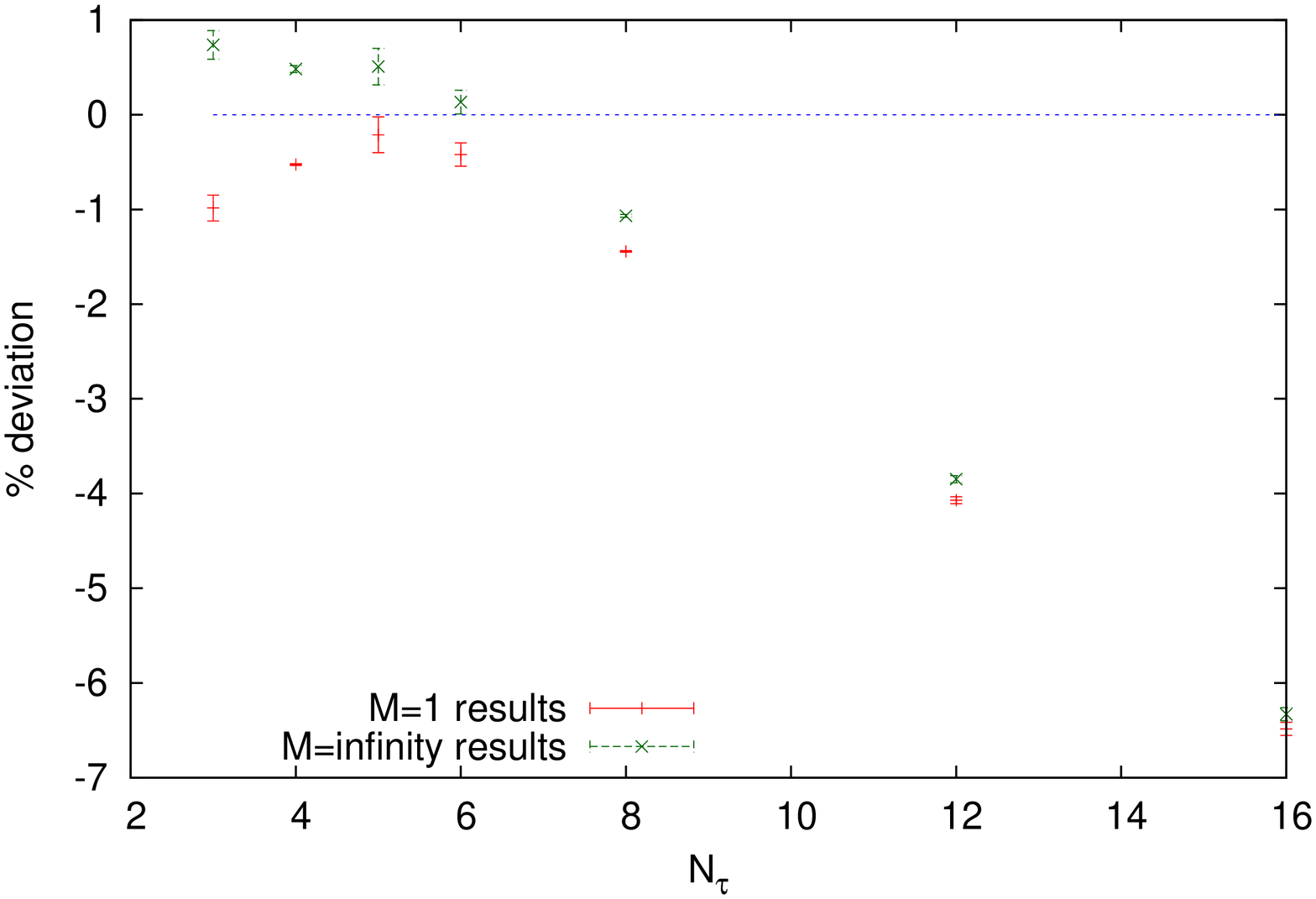}	
	\hspace{0.4cm}
 		\includegraphics[height=6.2cm,angle=270]{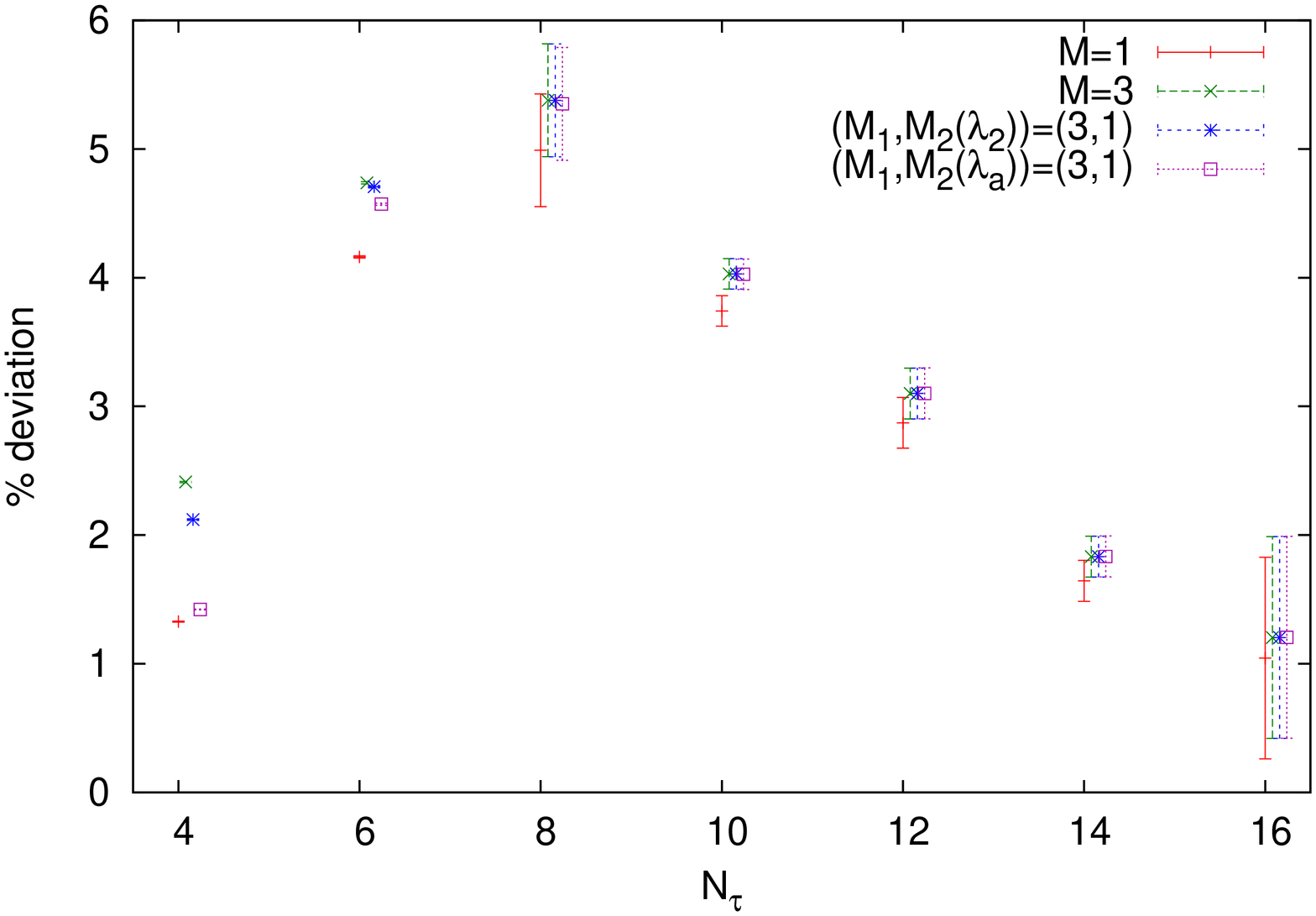}
	\caption[]{Relative error of $\beta_c$ predicted by the effective theories when compared 
	to simulations of the 4d theories, for $SU(2)$ (left) and $SU(3)$ (right).}
	\label{fig:betas}
\end{center}
\end{figure}

\begin{table}[t]
\begin{center}
\begin{tabular}{|c||c|c||c|}
\hline
	$N_\tau$ & $M=1$ & $M=\infty$ & $\mbox{4d YM}$ \\
\hline
	3    &    2.15537(89)   &     2.1929(13)   &     2.1768(30) \\
	4    &    2.28700(55)   &     2.3102(08)   &     2.2991(02) \\
	5    &    2.36758(40)   &     2.3847(06)   &     2.3726(45) \\
	6    &    2.41629(32)   &     2.4297(05)   &     2.4265(30) \\
	8    &    2.47419(22)   &     2.4836(03)   &     2.5104(02) \\
	12   &    2.52821(14)   &     2.5341(02)   &     2.6355(10) \\
	16   &    2.55390(10)   &     2.5582(02)   &     2.7310(20) \\
\hline
\end{tabular}
\caption{Critical couplings $\beta_c$  for $SU(2)$ from two effective 
theories compared to simulations of the 4d theory \cite{fingberg_heller_karsch_1993,bo,Ve}).}
\label{tab:su2_betas}
\end{center}
\end{table}~\begin{table}[t]
\begin{center}
\begin{tabular}{|c||c|c|c|c||c|}
\hline
	$N_\tau$ & $M=1$ & $M=3$ & $M_1,M_2(\lambda_2)=3,1$ &
	$M_1,M_2(\lambda_a)=3,1$ & $\mbox{4d YM}$\\
\hline
	4	&  5.768 & 5.830 & 5.813 & 5.773 &5.6925(002) \\
	6	&  6.139 & 6.173 & 6.172 & 6.164 &5.8941(005) \\
	8	&  6.300 & 6.324 & 6.324 & 6.322 &6.0010(250) \\
	10	&  6.390 & 6.408 & 6.408 & 6.408 &6.1600(070) \\
	12	&  6.448 & 6.462 & 6.462 & 6.462 &6.2680(120) \\
	14	&  6.488 & 6.500 & 6.500 & 6.500 &6.3830(100) \\
	16	&  6.517 & 6.528 & 6.528 & 6.528 &6.4500(500) \\
\hline
\end{tabular}
\caption{Critical couplings $\beta_c$  for $SU(3)$ from different effective 
theories compared to simulations of the 4d theory \cite{fingberg_heller_karsch_1993, Kogut_et_al_1983}).}
\label{tab:su3_betas}
\end{center}
\end{table}
The agreement is remarkable in all cases, with the relative error of the effective theory 
results compared to the full ones shown in Fig.~\ref{fig:betas}. 
The comparison of alternative truncations of the logarithm shows once more that it has almost no 
influence on the accuracy of the final result, as described earlier.
Interestingly, there appears to be
a `region of best agreement', with the deviation growing both for small and large $N_\tau$.
We ascribe this to the fact that there are two competing systematic errors, as discussed earlier:
the validity of the strong coupling series for a given coupling $\lambda_i$ is better
the smaller $\beta$ and hence $N_\tau$, whereas the truncation of the next-to-nearest neighbour
interactions gains validity with growing $N_\tau$. In particular in the case of $SU(3)$, there appears
to be a cancellation of the two kinds of systematics, rendering the effective description
better for the original theory on finer lattices.

The strong-coupling series was inspected both by comparing
the resulting $\beta_c$ from series of different depth and by Pad\'e analysis,
and we observe a satisfactory convergence.
It was also found that the error due to the truncation of the
strong-coupling series is much larger than that from neglecting higher couplings.

One can also compare the results presented here with those from the inverse Monte Carlo approach,
where the effective theory is found in a completely non-perturbative way; inspection of the $SU(2)$
case \cite{Heinzl:2005xv}, in particular, shows that the abrupt change of curvature in the inverse Monte Carlo
function $\lambda_1(\beta)$ at the critical point is not captured by our strong-coupling approach,
which is consistent with $\beta_c$ marking the radius of 
convergence also for the series expansion of the effective coupling $\lambda_1$. Thus, the
inverse Monte Carlo approach has a wider range of validity whereas the series approach
furnishes analytically known mappings between the full and effective theories.
 
\section{Conclusions}
\label{sec:conclusions}

We have derived, by means of strong coupling expansions, an effective description for lattice
pure gauge theories at finite temperature which respects explicitly the requirement of centre
symmetry and has only scalar Polyakov loop variables as degrees of freedom. Moreover,
due to the dimensional
reduction involved, the $N_\tau$-dependence is encoded completely in the maps from the effective to
the original coupling $\beta$, whose expansion can be extended, in principle, to higher orders.
We have also considered interaction terms other than the leading one, namely a next-to-nearest
neighbour interaction and an adjoint-representation coupling term.

Our Monte Carlo approach to the models, while requiring modest computational resources,
confirms the expected nature of the symmetry-breaking transition for both $SU(2)$- and
$SU(3)$-based effective formulations, and allowed us to predict the critical
point $\beta_c$ of the original 4d thermal gauge theories with an accuracy
within a few percent for a variety of values of $N_\tau$. Particular attention was
devoted to estimating the effect of employing different approximations, 
with quite stable answers in support of the good convergence of the series,
and of neglecting higher-order interaction terms, which again does not have 
a strong effect on the final answers especially at finer lattices.

An extension of the present work could be the study of $SU(N)$ gauge theories with $N>3$
(cf.~\cite{largen} and references therein), which can be performed much in the same way as 
the cases examined here; even more intriguing is the possibility to keep 
the theory simple while getting a step closer to physical QCD, i.~e.~by introducing
fermions and finite baryon density, for instance by employing a hopping parameter 
expansion \cite{Langelage:2009jb,Langelage:2010yn}.

\section*{Acknowledgements}
S.~L. and O.~P. are partially supported by the German BMBF grant \textit{FAIR theory: the QCD
phase diagram at vanishing and finite baryon density},  06MS9150, and by
the Helmholtz International Center for FAIR within the LOEWE program of the State of Hesse. 
J.~L.~acknowledges financial support by the EU project \textit{Study of Strongly interacting 
Matter}, No. 227431, and by the BMBF under the project \textit{Heavy Quarks as
 a Bridge between Heavy Ion Collisions and QCD}, 06BI9002.

\end{document}